# Pulsed Laser Deposition of a PBN:65 Morphotropic Phase Boundary Thin Film with Large Electrostriction

Xi Yang [1], Andrew Beckwith [2], and Mikhail Strikovski [2]

1) Fermi National Accelerator Laboratory, P.O. Box 500, Batavia, Illinois 60510-0500

2) Department of Physics, University of Houston, 77204-5005

*Abstract*

We deposited epitaxial thin films of Morphotropic Phase Boundary (MPB) $Pb_{0.65}Ba_{0.35}Nb_2O_6$ (PBN:65) on MgO substrates using pulsed laser deposition. Afterwards, a novel transmission optical experiment was developed to measure the electric field-induced bending angle of the thin film sample using a divergent incident light. From which the electric field-induced strain was obtained, and it was used to calculate the electrostrictive constant of the PBN thin film. The result is 0.000875 $\mu m^2/V^2$, and it is consistent with what we measured in the reflection experiment.

I. INTRODUCTION

The ferroelectric lead barium niobate solid-solution system, $Pb_{1-x}Ba_xNb_2O_6$ (PBN:1-x), is potentially important for optical and electronic applications because of its unique properties as a tungsten bronze-type ferroelectric relaxor with a MPB that separates the tetragonal ferroelectric phase from the orthorhombic ferroelectric phase [1]-[5]. For PBN compositions near the MPB, the electro-optic (EO) [6], piezoelectric [3], electrostrictive, and pyroelectric [7] properties are enhanced and are expected to be temperature invariant because the MPB composition does not vary significantly with temperature. Furthermore, the large spontaneous polarization and high switchable polarization near this boundary are especially significant for guided-wave electro-optic devices [8], [9], surface acoustic wave (SAW) devices, sensors, actuators and adaptive structures.

Due to the loss of $Pb^{2+}$ during the growth, very limited work has been done so far on PBN since its discovery in 1960. The production of high quality thin films not only provides a possible solution in this situation but also has been driven by the trend toward miniaturization of electronic components, replacement of expensive single-crystal ferroelectrics, and the flexibility of thin films for integration into different types of devices [10]. Among thin-film growth techniques, such as plasma sputtering, ion beam sputtering, metal organic chemical vapor deposition (MOCVD), sol-gel spin coating, metal organic decomposition (MOD), and pulsed laser deposition (PLD), PLD has a number of advantages over the other techniques because ablation and deposition rates are readily controllable, ablation occurs stoichometrically, and few parameters need adjustment during the growth. So we chose PLD as the means of growing high quality PBN:65 thin films.

## II. GROWTH AND CHARACTERIZATION OF PBN FILMS

First, PBN:65 films were deposited on MgO substrates using PLD technique due to the consideration of lattice matching between thin films and substrates. Here, a lattice constant of a = 12.446 Å [11], PBN:65 grows nearly lattice matched within 1.56% to 3 unit cells of MgO (a = 4.213 Å). Deposition was carried out under the conditions of 850 mJ per laser pulse at 5 Hz, a 700°C substrate temperature and a 100-mTorr oxygen partial pressure. The X-ray $2\theta$ scan of the PBN:65 film on (001) MgO substrate is shown in Fig. 1. The intensity profiles of (001) and (002) peaks from the film clearly indicate the epitaxial growth of the c-direction oriented film, and their locations are used to calculate the lattice constant of c = 3.93±0.006Å.

Measuring the piezoelectric and electrostrictive coefficients of a thin film requires the application of an electric field along the in-plane directions. Planar interdigital electrodes (IDE) consisting of 10-μm conducting lines separated by 10-μm gaps were deposited on top of the PBN thin film to apply an in-plane electric field. Besides that, IDE are appropriate for obtaining large electric fields even at relatively low voltages due to the small space between two neighboring conducting lines. Each line consists of a thin layer of chromium (2500 Å), and external contacts were made to these electrodes using soft indium. When a DC voltage is applied to the contacts on the sample, a spatial modulation of the strain S occurs in the film along the in-plane electric field direction. This strain modulation causes a shear force, and this force acts on the MgO substrate and causes it to bend. The bending direction depends on the properties of the shear force. A contractive force causes the substrate to bend toward the film whereas an expansion force causes the substrate to bend away from the film. In our situation, the field-induced expansion force causes the MgO substrate to bend away from the film and this will be mentioned in the following experiment.

In our procedures, we measure the electric field-induced strain by measuring the electric field-induced bending of the PBN:65 thin film sample. Afterwards, we use a bending model developed in our lab to extract the strain from the bending measurement. In the electric field-induced bending measurement, we did both the transmission and reflection experiments using a 1 mW He-Ne laser operating at the wavelength of 632.8 nm. In the transmission experiment, a beam with a 1° divergence was used to illuminate the PBN sample. See Fig. 2 for details of the experimental setup. A DC power supply was used to apply the electric field to the sample. A CCD camera was used to take photographs of the diffraction pattern, which was the superposition of both the grating images (with a small period) and the interference pattern (with a large period) from the MgO substrate [12]. From the images taken by the CCD camera after the electric field was applied to the PBN thin film sample, the interference pattern from the MgO substrate was moving across the grating images when the electric field-induced sample bending happened. A major advantage about the appearance of the grating images in the diffraction pattern was that the grating images could be used as a built-in scale to measure the shift of the interference pattern caused by the field-induced bending of the MgO substrate. See Fig. 3 for the details. We used a point source approximation to represent the incident beam with a 1°-divergence. Small angle approximation was used to calculate the bending angle θ, as shown in the equation 1.

$$\theta \approx \frac{\Delta N \cdot (a+b)}{d}. \tag{1}$$

Here, $\Delta N$ was the number of grating lines passed by one of the interference peaks from the MgO substrate when 20 volts were applied to the sample. $d$ was the distance between the point source and the grating, $a$ was the grating space, and $b$ was the width of the grating line. In our experiment, $a$ was 10 μm, $b$ was 10 μm, and $d$ was 22.0±0.1 mm. From the 3-D diagram, as shown in Fig. 3, we knew that $\Delta N$ was approximately 15.3±0.2 and it was measured after the bending stopped. The precision of $\Delta N$ decided the precision of the bending angle $\theta$. Since $\Delta N$ was obtained by counting the number of grating lines, which were passed by one of the interference peaks from the MgO substrate when the electric field was applied, the accuracy of defining an interference peak determined the precision of $\Delta N$, and the error limited by the contrast of the interference pattern was estimated to be less than one fifth of a grating period (a+b). 150 images were taken by the CCD camera, shown in Fig. 2, at a speed of one image per second when 20 volts were applied to the thin film sample. Afterwards, each image was integrated along the direction of the grating line, and the result was used as a time slit of the total 150 seconds in Fig. 3. Using Eq. (1), we obtained the bending angle $\theta$ of the MgO substrate, and it was 0.80°±0.04°. Since the bending angle was obtained after the electric field-induced bending stopped, it was the maximum bending angle that we could get when 20 volts were applied to the contacts on the sample. All the bending angles were measured in the same condition and we won't mention it later.

We develop a bending model in order to calculate the strain of the MgO substrate when the bending angle is known. See Fig. 4 for details. The strain $S$ is defined as the ratio of the deformation $\Delta x$ over the region $l$ where the deformation occurs, and the small angle approximation is used since the bending angle is less than 1°. The strain is expressed as

$$S = \Delta x / l \approx t \cdot \theta / (2 \cdot l). \qquad (2)$$

Here t was the thickness of the MgO substrate (1 mm). $l$ was the length of the grating region, over which the electric field can be applied, and it was 2 mm. The strain of the film was equal to the strain of the MgO substrate, which was $S = 0.0035±0.0002$. Since a quadratic relationship between the stain $S$ and the applied electric field $V$ made a good fit to the experimental data, as shown in Fig. 5, and also the bending angle stayed the same when the applied electric field was reversed in the experiment, we treated the bending phenomenon as the result of electrostriction. It is likely that the quadratic relation between the strain and the applied electric field is caused by the in-plane polycrystalline structure of the thin film according to the X-ray in-plane measurement. However, this needs some further investigations, such as by improving the in-plane structure of the film to be closer to a single-crystalline, whether the relationship between the strain and the applied electric field will become linear or not. According to the IEEE standard [13] (adapted for electrostriction), the following equation is used to calculate the effective electrostrictive constant $M_{pij}$

$$S_{pij} = M_{pij} E_i E_j. \tag{3}$$

The electric field was calculated using the equation

$$E \approx \frac{V}{a}. \tag{4}$$

Therefore the electrostrictive constant of the PBN film was $0.000875 \pm 0.00005$ $\mu m^2/V^2$. We knew that the film was always under stress from the MgO substrate once the field was applied. After the bending stopped, the stress in the film can be calculated from the stress in the MgO substrate because they reached equilibrium. The equation

$$T \approx Y \cdot S \tag{5}$$

was used to calculate the stress [13], where $Y = 3.156 \times 10^7$ (N/cm$^2$) [14] (Young's modulus of MgO). Because the strain $S$ was $0.0035 \pm 0.0002$, the stress $T$ was $(1.10 \pm 0.06) \times 10^5$ (N/cm$^2$). Unfortunately, because Young's modulus of the film is unknown, the electrostrictive constant of the film in zero stress cannot be calculated. So in our case, the electrostrictive constant of the PBN:65 film is the value under $1.10 \times 10^5$ (N/cm$^2$) stress. It is true that the electrostrictive constant in zero stress should be larger than 0.000875±0.00005 $\mu m^2/V^2$. Here, since the in-plane structure of the thin film is polycrystalline, the electrostrictive constant is an effective value. We called it $M_{eff}$.

We designed a reflection experiment in order to simplify the experimental procedures used for the electric field-induced bending measurement in the transmission experiment and also examine the result obtained in the transmission experiment. The details about the reflection experiment are in [12]. From the reflection experiment we obtained the electrostrictive constant, which was 0.000875±0.00005 $\mu m^2/V^2$ and the same with what we obtained in the transmission experiment.

## III. CONCLUSIONS

We have deposited high-quality epitaxial PBN:65 thin films on MgO substrates using PLD. The effective electrostrictive constant of these films was measured in both the transmission and reflection experiments, and they were consistent with each other. In order to make an approximated comparison with other MPB ferroelectric thin films, we converted the effective electrostrictive constant $M_{eff}$ of the PBN:65 thin film to an effective piezoelectric constant $d_{eff}$ using equation

$$d_{eff} = M_{eff} \cdot E. \qquad (6)$$

The result was 1750 pm/V, and it was much larger than that of 844 pm/V, which was reported from a SBN:75 thin film sample [15], [16].

The optical techniques used in our experiment can be generally applied to characterize the piezoelectricity and electrostriction of ferroelectric thin films. The large electrostrictive constant obtained from the PBN:65 thin film indicates that the PBN:65 thin films can be potentially used to improve device performance in surface acoustic wave (SAW), adaptive optics and microelectromechanical structures (MEMS), etc. For example, one expects that the applied electric field to the PBN:65 thin film sample could be further increased either by decreasing the grating space or by increasing the applied DC voltage, and the surface curvature of the thin film sample could be varied in the range of several degrees under a moderately applied DC voltage. It is possible for PBN:65 thin films to be integrated into adaptive optics with capabilities of varying the focal length and redirecting light through the voltage control.

The authors would like to thank Dr. Leiming Xie for useful discussions about the mechanical properties of the thin films and Dr. Lowell T. Wood for providing the optical instruments for the thin film characterization.

**References**

[1] T. R. Shrout, L. E. Cross, and D. A. Hukin, "Ferroelectric properties of tungsten bronze barium niobate (PBN) single crystals," *Ferroelectric Lett.*, vol. 44, no. 11, pp. 325-330, 1983.

[2] M. Adachi, S. G. Sankar, A. S. Bhalla, Z. P. Chang, and L. E. Cross, "Growth and dielectric properties of lead barium niobate single crystals and morphotropic phase boundary," in *Proc. 6th IEEE Int. Symp. Appl. Ferroelectrics*, Bethlehem. PA, pp. 169-.171, 1986.

[3] T. R. Shrout, H. Chen, and L. E. Cross, "Dielectric and piezoelectric properties of $Pb_{1-x}Ba_xNb_2O_6$ ferroelectric tungsten bronze crystals," *Ferroelectrics*, vol. 74, no. 3-4, pp. 317-324, 1987.

[4] J. R. Oliver and R. R. Neurgaonkar, "Ferroelectric properties of tungsten bronze morphotropic phase boundary systems," *J. Am. Ceram. Soc.*, vol. 72, no. 2, pp. 202-211, 1989.

[5] L. E. Cross, "Relaxor ferroelectrics," *Ferroelectrics*, vol. 76, no. 3-4, pp. 241-267, 1987.

[6] M. Didomenico, Jr. and S. H. Wemple, "Oxygen-octahedra ferroelectrics I. Theory of electro-optical and nonlinear optical effects," *J. Appl. Phys.*, vol. 40, no. 2, pp. 720-734, 1969.

[7] R. Lane, D. L. Mack, and K. R. Brown, "Dielectric, piezoelectric and pyroelectric properties of the $PbNb_2O_6$-$BaNb_2O_6$ system," *Trans. J. Brit. Ceram. Soc.*, vol. 71, no. 1, pp. 11-22, 1972.

[8] H. F. Taylor, "Application of guided-wave optics in signal processing and sensing," *Proc. IEEE*, vol. 75, no. 11, pp. 1524-1535, 1987.

[9] L. Thylen, "Integrated optics in LiNbO3: recent developments in devices for telecommunications," *J. Lightwave Technology*, vol. 6, no. 6, pp. 847-861, 1988.

[10] Y. Xu, *Ferroelectric Materials and Their Applications*. North-Holland, New York, 1991, pp. 206-210.

[11] M. Lee and R. S. Feigelson, "Growth of lead barium niobate (Pb1-xBaxNb2O6) crystals by the vertical Bridgman method II. $Sr_{0.61}Ba_{0.39}Nb_2O_6$-seeded growth," *J. Crystal Growth*, vol. 193, pp. 355-363.

[12] X. Yang, *Studies of Ferroelectric Materials Using Novel Optical Techniques*. Ph.D. Dissertation, University of Houston, 2001, pp. 64-125.

[13] C. Z. Rosen, B. V. Hiremath, and R. Newnham, *Key Papers in Physics Piezoelectricity*. AIP, pp. 235-280, 1992.

[14] G. Simmons and H. Wang, *Single Crystal Elastic Constants and Calculated Aggregrate Properties: A Handbook*. M. I. T. press, pp. 214, 1971.

[15] P. Tayebati, D. Trivedi, and M. Tabet, "Pulsed laser deposition of SBN:75 thin films with electro-optic coefficient of 844 pm/V", *Appl. Phys. Lett.*, vol. 69, pp. 1023-1025, 1996.

[16] D. A. Mendels, *An Introduction to Modeling of Piezoelectric Thin Films*. http://www.npl.co.uk/materials/functional/thin_film/modelling/func_modelling.html.

**Figure Captions**

Fig. 1. X-ray 2θ scan of $Pb_{0.65}Ba_{0.35}Nb_2O_6$ film grown on MgO substrate.

Fig. 2. Schematic diagram of the experimental setup for the electrostriction measurement of the PBN thin film. ODF represents optical density filter.

Fig. 3. In this intensity contour *vs*. position and time, the dashed lines represent the grating images, which are nearly stationary with time. The black structures, which are changing with time, represent the interference from the MgO substrate. In this diagram, we can use the number of the grating images (dashed lines) passed by each of the interference peaks (black structures) to estimate the field-induced bending angle by Eq.(1).

Fig. 4. Schematic illustration of the bending process. Bending occurs when an electric field is applied to the PBN thin film, which is on the top of the MgO substrate. The rectangle with mid-line (dashed lines) represents the MgO substrate before bending. Here, the mid-line represents the neutral plane. The curved plate represents the MgO substrate after bending.

Fig. 5. Fitting a quadratic relationship of $S=aV^2+bV+c$ to the experimental data of the saturated strain *S vs*. the applied DC voltage *V*. Here, $a=9\times10^{-6}$, $b=-2\times10^{-6}$, $c=-5\times10^{-7}$.

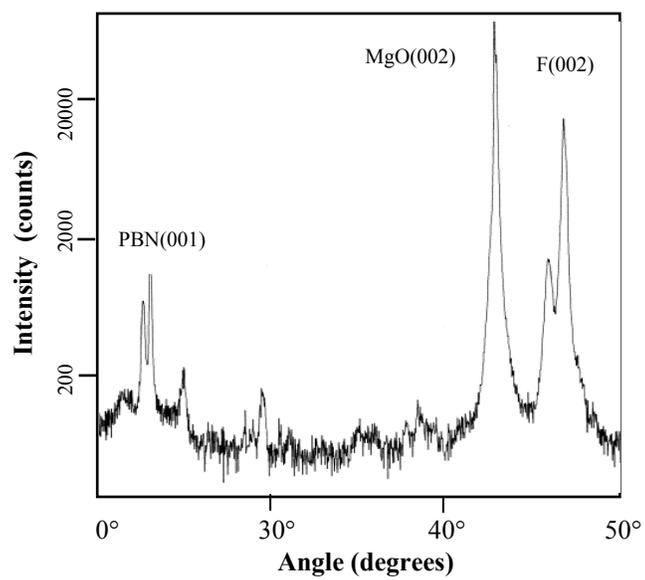

Fig. 1

Yang, et al.

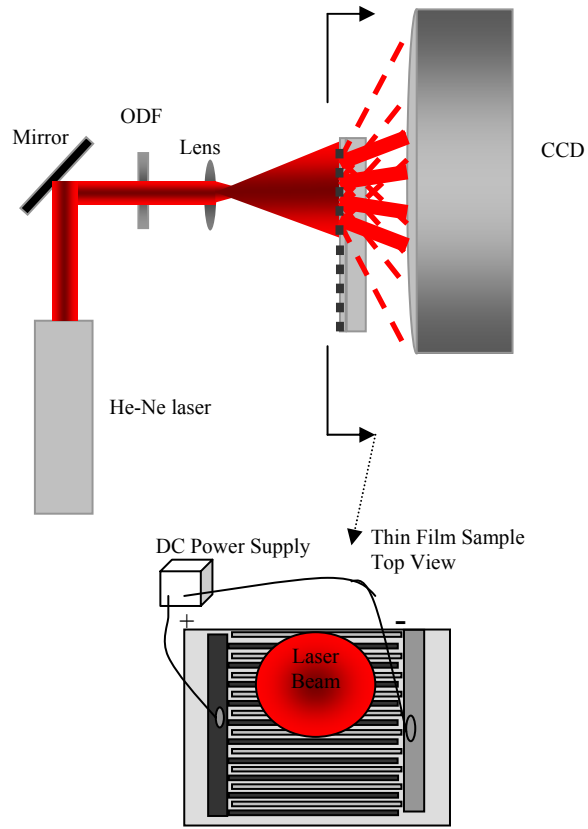

Fig. 2

Yang, et al

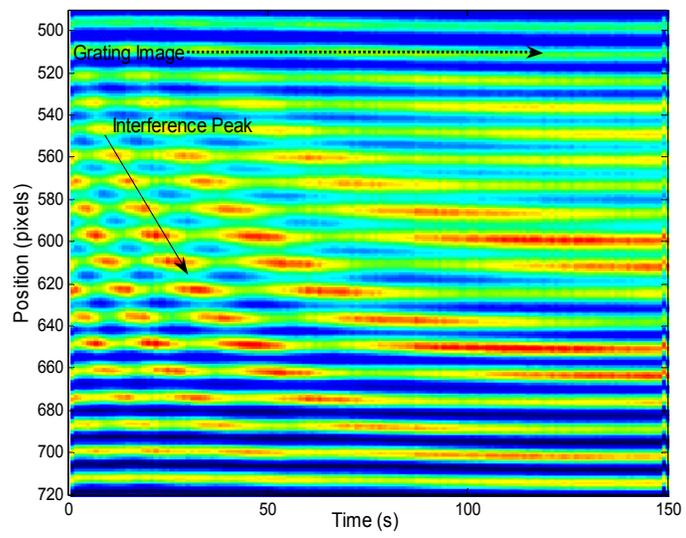

Fig. 3

Yang, et al

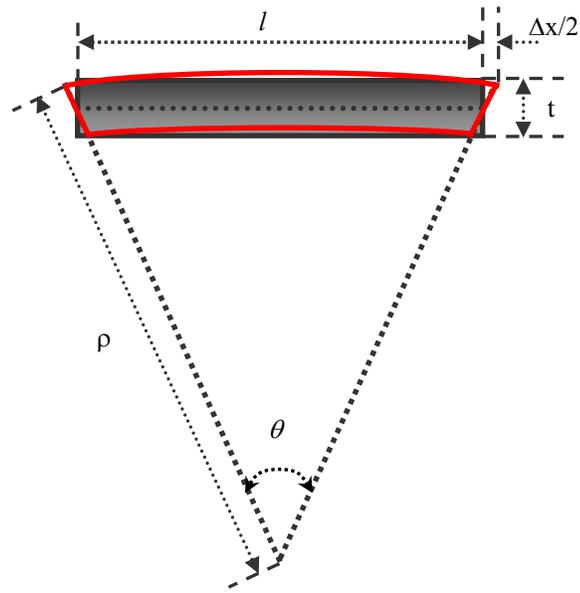

Fig. 4

Yang, et al.

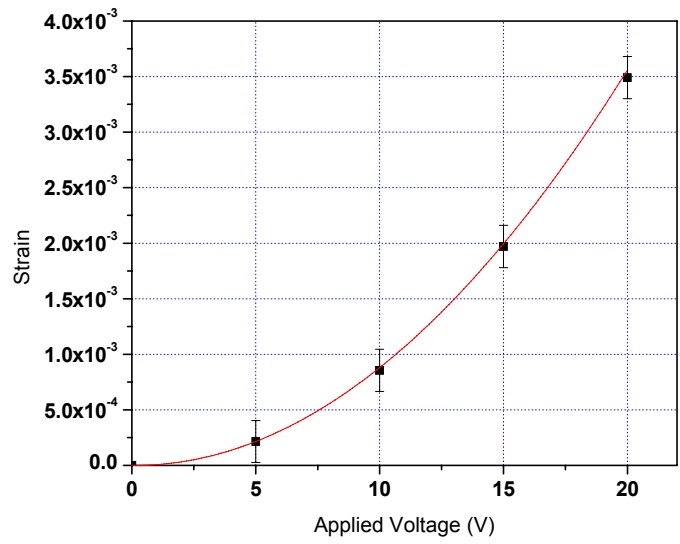

Fig. 5

Yang, et al.